\documentclass{article}

\usepackage{microtype}
\usepackage[final]{graphicx}
\usepackage{epstopdf}
\usepackage{subfigure}

\usepackage{booktabs} 
\graphicspath{ {./images/} }
\usepackage{hyperref}

\usepackage[accepted]{icml2019}

\icmltitlerunning{Unsupervised Representation Learning of DNA Sequences}

\begin{document}

\twocolumn[\icmltitle{Unsupervised Representation Learning of DNA Sequences}




\begin{icmlauthorlist}
\icmlauthor{Vishal Agarwal}{eee}
\icmlauthor{N. Jayanth Kumar Reddy}{eee}
\icmlauthor{Ashish Anand}{cse}
\end{icmlauthorlist}

\icmlaffiliation{eee}{Department of Electronics and Electrical Engineering, Indian Institute of Technology Guwahati, India}
\icmlaffiliation{cse}{Department of Computer Science and Engineering, Indian Institute of Technology Guwahati, India}

\icmlcorrespondingauthor{Vishal Agarwal}{vishalagarwal.jss@gmail.com}

\icmlkeywords{Machine Learning, ICML}

\vskip 0.3in
]



\printAffiliationsAndNotice{}  

\begin{abstract}
Recently several deep learning models have been used for DNA sequence based classification tasks. Often such tasks require long and variable length DNA sequences in the input. In this work, we use a sequence-to-sequence autoencoder model to learn a latent representation of a fixed dimension for long and variable length DNA sequences in an unsupervised manner. We evaluate both quantitatively and qualitatively the learned latent representation for a supervised task of splice site classification. The quantitative evaluation is done under two different settings. Our experiments show that these representations can be used as features or priors in closely related tasks such as splice site classification. Further, in our qualitative analysis, we use a model attribution technique \textit{Integrated Gradients} to infer significant sequence signatures influencing the classification accuracy. We show the  identified splice signatures resemble well with the existing knowledge.
\end{abstract}

\section{Introduction}
\label{introduction}

Recently there is a surge in studies using deep learning models for DNA sequence based classification tasks. One of the primary reason for the adoption of such methods is representation learning or feature learning from raw data. In the case of DNA sequence based classification tasks, DNA sequence containing 4 nucleotides A, T, G, C constitute raw data. Most studies often choose fixed-length DNA sequences as input by choosing a context window. However, in many cases, important nucleotides may not lie within the same context window size in all input sequences. Hence, there is a requirement of models which can handle long as well as variable length DNA sequences as inputs. Such a model can then take into account of both short (local) and long (global) range dependencies.

In this work, we primarily focus on learning representation for long, variable length DNA sequences. We use an autoencoder-based sequence-to-sequence LSTM model to learn representations by encoding the input sequence in a fixed-length latent embedding and then reconstruct back the original input sequence using just the embedding. The representations learned include model parameters and the fixed-length latent embedding. This allows the model to aggregate information by implicit learning its parameters as features which summarizes the input sequence well as a fixed-dimensional latent representation. The learned representations can then be used as features or apriori information in various tasks related to DNA sequences such as splice site prediction.



We evaluate our model on splice site classification task. In genomics, splicing is an important phenomenon, leading to protein diversity in the body. We performed two quantitative and a qualitative evaluation of the learned latent representation of input DNA sequence. The quantitative evaluation is carried in two different settings. For qualitative analysis, we use Integrated Gradients, a model attribution technique  proposed by \cite{intgrad}. This provides attribution of input feature to the predicted classification score and identify relevant region and motifs influencing splicing.


\begin{figure*}[ht]
\vskip 0.2in
\begin{center}
\centerline{\includegraphics[width=\linewidth]{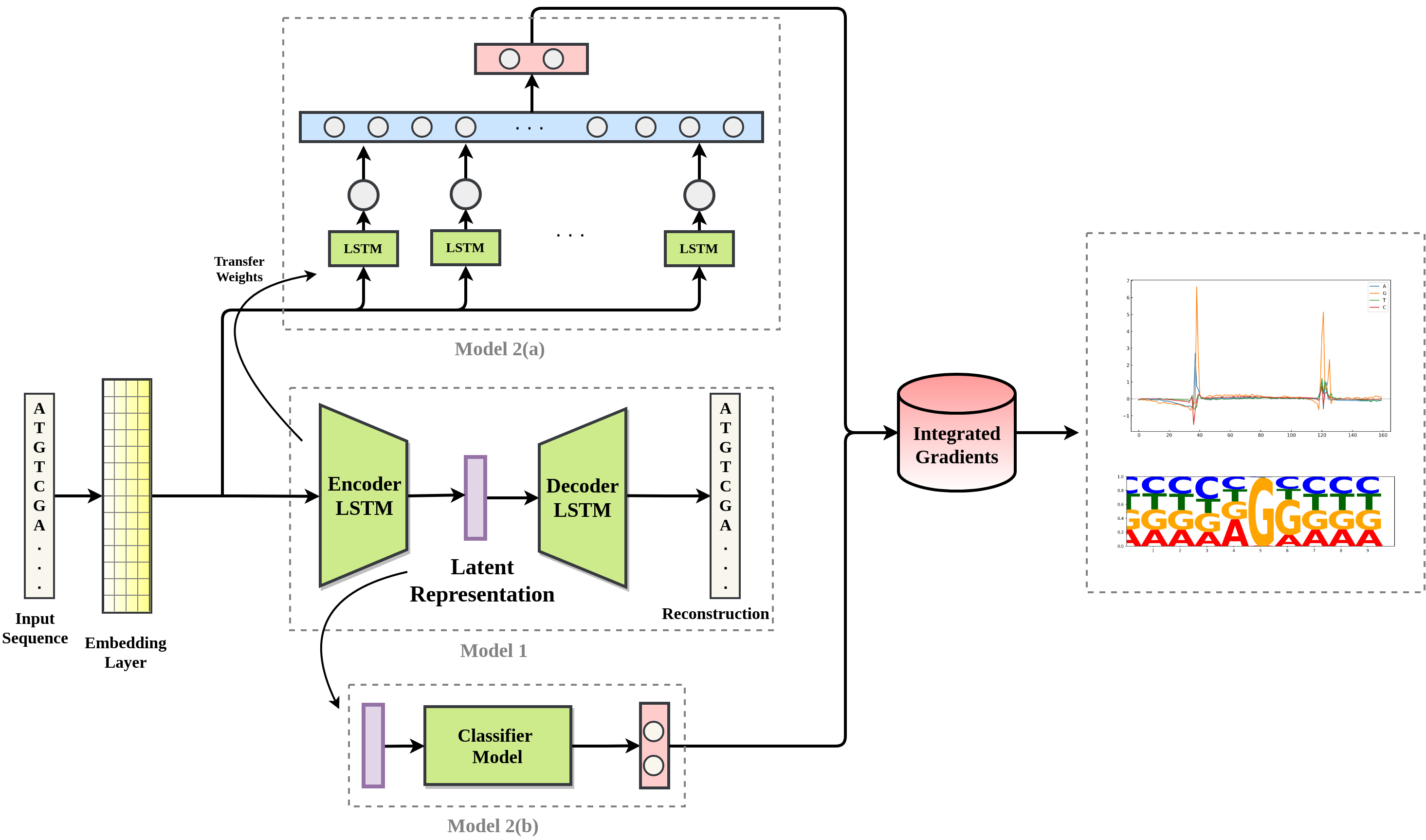}}
\caption{Graphical Illustration of the entire work. Model 1 corresponds to the sequence-to-sequence autoencoder. Model 2(a) and 2(b) shows the classifier for evaluation of learned representation on a supervised task. Integrated Gradient uses the prediction score to provide model attribution and aid in visualization.}
\label{workgraph}
\end{center}
\vskip -0.3in
\end{figure*}

\section{Related work}

Most of the previous work for feature learning and motif identification has been in the context of supervised learning. \cite{splicevec} proposed a method for learning distributed feature representation of splice junctions using k-mers. Spline transformation was introduced by \cite{spline} to improve upon traditional neural networks to learn better representations and improve prediction accuracy. \cite{mmn} used a memory matching network to dynamically learn a memory bank of motifs using sequence classification task. \cite{rbm} proposed a Restricted Boltzmann Machine based model with a new training method called boosted Contrastive Divergence to predict non-canonical splice sites and learn non-canonical feature vectors that couldn't be identified by traditional methods. Some works have also used Convolutional Neural Networks(CNN) in order to learn representations implicitly \cite{rover,deepbind, deepslice,deepmotif,aman}. 

Previous work leverages supervised labeled data to learn feature representation and motifs. Also, the features learned are task specific for a classification task in most cases. As per our knowledge, no work has been done on learning representations in an unsupervised setting for general use case. In our work, we try to learn representations in a more general unsupervised setting so that they are not task specific and can be used as features or priors in problems related to genomics. 


\section{Model Description}
In this section, we give a description of our model to learn fixed-length latent embedding of sequences. Figure \ref{workgraph} shows a complete graphical overview of the work.

\subsection{Sequence Autoencoder}
We use an autoencoder-like sequence-to-sequence model to learn fixed-length representations of sequences. The model (Model 1 in Figure~\ref{workgraph}) consists of an encoder and a decoder LSTM inspired by \cite{seq2seq}. The encoder network uses a bidirectional LSTM to process the input sequence from both ends and map it to a fixed-length embedding. Bidirectional LSTMs are used since it understands the context better by processing the input from both ends, from past to future and from future to past, and maintaining two hidden states. The encoder summarizes the input sequence and captures important motif information. The decoder network uses a unidirectional LSTM to reconstruct the input sequence using the latent embedding only. The motivation behind this is to capture relevant features that summarize the input sequence well-enough to be able to reconstruct it back.


\section{Experiment Design}

\subsection{Dataset}
We use GENCODE annotations \cite{gencode} based on human genome data GRCh38 to prepare the dataset. The model is trained on an earlier release version 20 and validated on a newer release version 26 for test after removing all common junction pairs present in version 20. Human genome data corresponding to 24 chromosomes namely 1 to 22, X and Y are used. Each of these chromosome consists of multiple genes and each gene is made up of multiple exons and introns. The intron length varied from 1 to about a million in length but we choose only those introns whose length is above 30. For true data or positive sequences, we choose intron sequences beginning with GT, corresponding to donor site (exon-intron boundary), and ending with AG, corresponding to an acceptor site (intron-exon boundary). This constitutes a splice site sample containing both donor and acceptor sites together. 
The junction pairs are extracted from protein-coding gene only. This left us with 290,502 positive samples from version 20 which we use for autoencoder training and 5,612 positive samples from version 26 which we use for testing or quantitative analysis discussed in section \ref{result}.

For negative data generation, we use a technique based on existing works \cite{deepslice,splicevec}. In this approach, false data is randomly sampled from the genome data based on some heuristics. For each false junction pair, the consensus dimer GT and AG is searched randomly such that both donor and acceptor are in the same chromosome and its distance is in the range of true data length range. We create a huge list of negative sequences and then sampled some sequences such that the number of negative samples is equal to the number of positive samples.

\subsection{Unsupervised Learning}
The autoencoder model takes DNA sequences as input and uses it in an unsupervised way without labels to learn latent representations. The one-hot representation is fed to an embedding layer which then goes into the encoder bi-LSTM. The encoder outputs a fixed-length 256-dimensional vector which is the latent embedding summarizing the input. The latent vector is then fed to the decoder LSTM to predict the actual input sequence at each time step. At the decoder side, the output is computed as softmax over nucleotide A, T, G, and C. 

The loss function criterion used was minimizing the cross-entropy. The model was trained over 300 epochs with Adam as the optimization algorithm. The entire experiment was performed on NVIDIA TitanXP GPU with 12GB memory. The hyperparameter of the model is the dimension of latent representation. We experiment with 128, 256 and 512 dimensional representation and found 256 to be performing best, giving the lowest cross-entropy loss.

The next section describes the quantitative and qualitative analysis of the learned representation. This is done in order to make sure that the representations learned are useful.

\section{Results and Discussion}
\label{result}
In this section, we describe the quantitative and qualitative analysis of the learned representation and discuss results for the same. 

\subsection{Splice site classification}
The quantitative analysis of learned representations is done on a supervised task under two different settings. First, an LSTM is trained for splice site prediction in a DNA sequence. Instead of initializing the LSTM with random weights, it is initialized with the trained encoder weights to add apriori information (Model 2a in Figure~\ref{workgraph}). This provides a good starting point for the discriminative model to converge faster and improves classification accuracy. We compare this model with a baseline model of similar architecture but randomly initialized parameters. Table~\ref{pretrain} shows the former model performs better. We also experimented with different architectures such as LSTM, bidirectional LSTM and bidirectional LSTM with Attention and compared the results. Table \ref{pretrain} shows the comparison of different types of models. 

In the second evaluation setting, the latent embeddings are used as features on the same task of splice site identification (Model 2b in Figure~\ref{workgraph}). The difference between the previous and this model is that the former uses the DNA sequence as input whereas the latter uses just 256-dimensional fixed length latent embedding as the input feature vector. We use Support Vector Machine(SVM), 2-layer Artificial Neural Network(ANN) and a vanilla Recurrent Neural Network(RNN) model to conclude the effectiveness of latent representations. If it did capture motif information, then we expect the classifier to perform well. The results for this setting are shown in table \ref{classifier}.

\begin{table}[h]
\caption{Classification accuracy for encoder initialized LSTM model}
\label{pretrain}
\vskip 0.15in
\begin{center}
\begin{small}
\begin{sc}
\begin{tabular}{llc}
\toprule
 & Model & Accuracy \\
\midrule
Random  & LSTM & 95.43\% \\
Weights & Bi-LSTM & 96.04\% \\
 & Bi-LSTM Attention & 97.23\% \\
\midrule
Autoencoder & LSTM & 98.54\% \\
Initialized & Bi-LSTM & 98.60\% \\
 & Bi-LSTM Attention & 99.07\% \\
\bottomrule
\end{tabular}
\end{sc}
\end{small}
\end{center}
\vskip -0.1in
\end{table}

\begin{table}[h]
\caption{Classification accuracy for simple classifier model}
\label{classifier}
\vskip 0.15in
\begin{center}
\begin{small}
\begin{sc}
\begin{tabular}{lc}
\toprule
Model & Accuracy \\
\midrule
SVM    & 98.63\% \\
ANN & 98.88\% \\
Vanilla RNN & 98.93\% \\
\bottomrule
\end{tabular}
\end{sc}
\end{small}
\end{center}
\vskip -0.1in
\end{table}

\subsection{Model Attribution}
This section describes the qualitative analysis of our model to provide attribution of the input sequence to the predicted output of the supervised task. To achieve this, we use a popular visualization technique - Integrated Gradients proposed by \cite{intgrad}. The visualizations provide model attribution by capturing important regions in the sequence. This helps us to identify motifs present in the sequence. These motifs can be interpreted as signals which influence splicing.
3
Integrated gradient requires a baseline against which it compares the prediction of the network and accordingly provides attribution to the feature which differs in the input and the baseline. In our case, we chose the baseline as zero embedding matrix. It calculates the attribution score by accumulating gradient of network prediction score with respect to the embedding at each point in the straight-line path from baseline to input, then multiplied by the difference in the baseline and input feature value. In our experiment, we used 200 steps for gradient calculation along the path.

\begin{figure}[t]
\vskip 0.2in
\begin{center}
\subfigure[Donor]{\label{totaldon}\includegraphics[width=40mm]{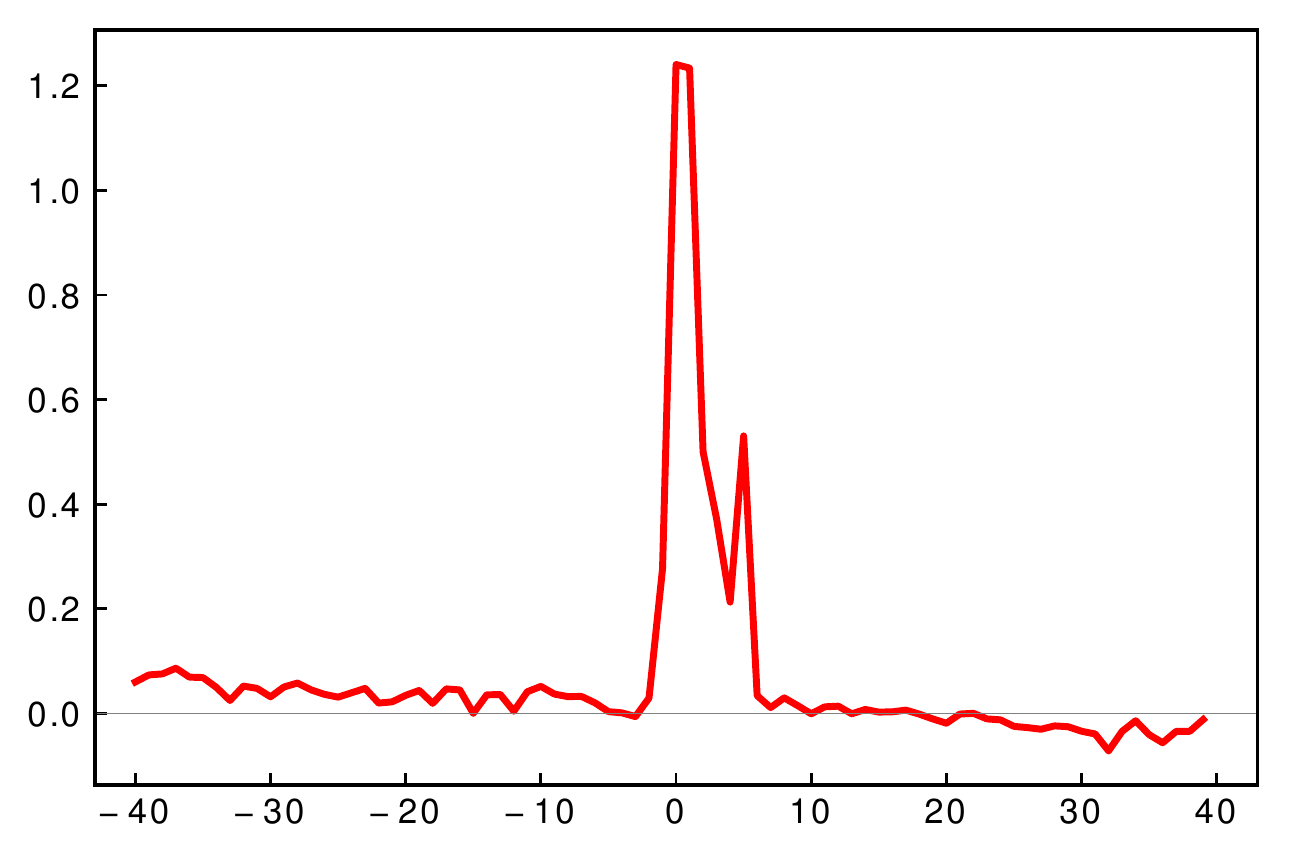}}
\subfigure[Acceptor]{\label{totalacc}\includegraphics[width=40mm]{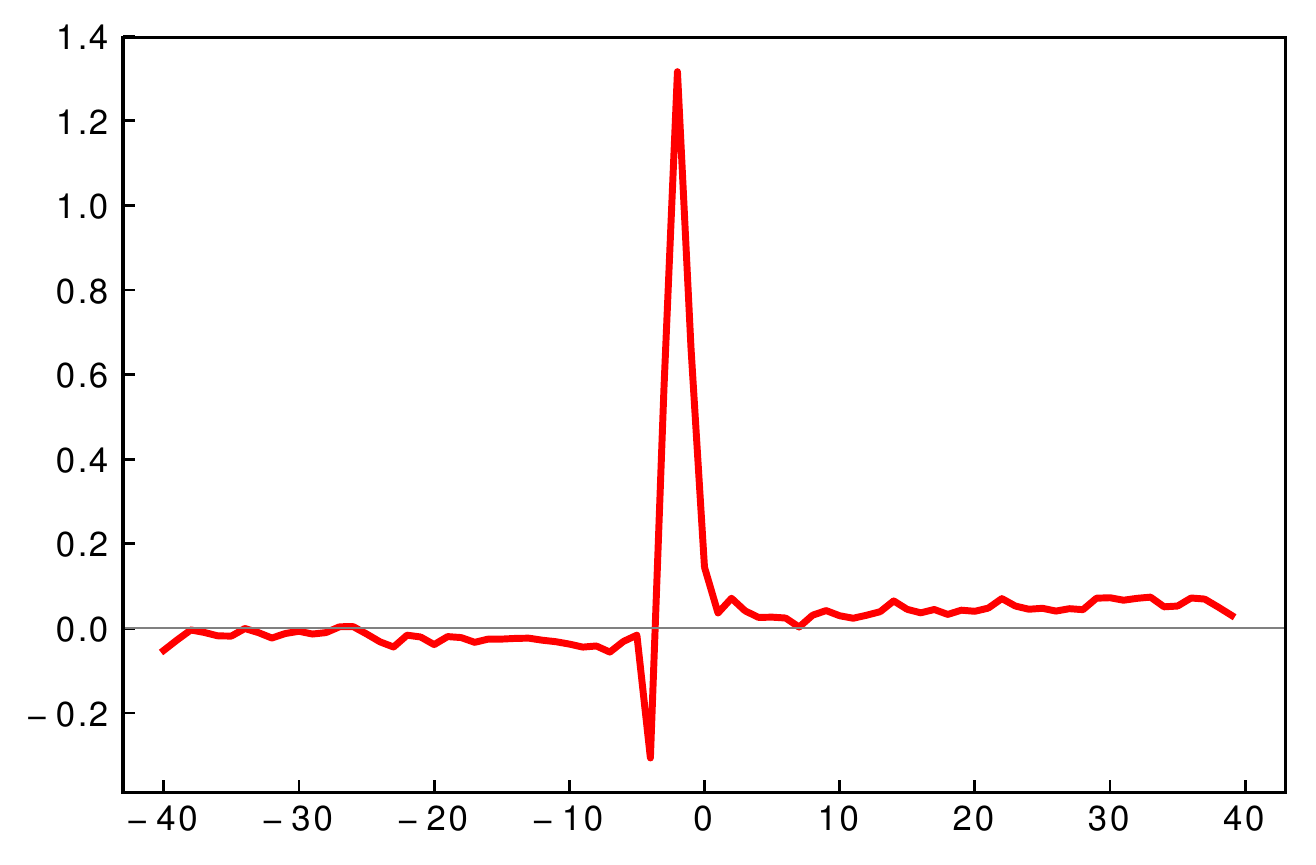}}
\caption{Average attribution score per position}
\label{totalscore}
\end{center}
\vskip -0.3in
\end{figure}

\begin{figure}[t]
\vskip 0.2in
\begin{center}
\label{pmscore}
\subfigure[Donor]{\label{pmdon}\includegraphics[width=40mm]{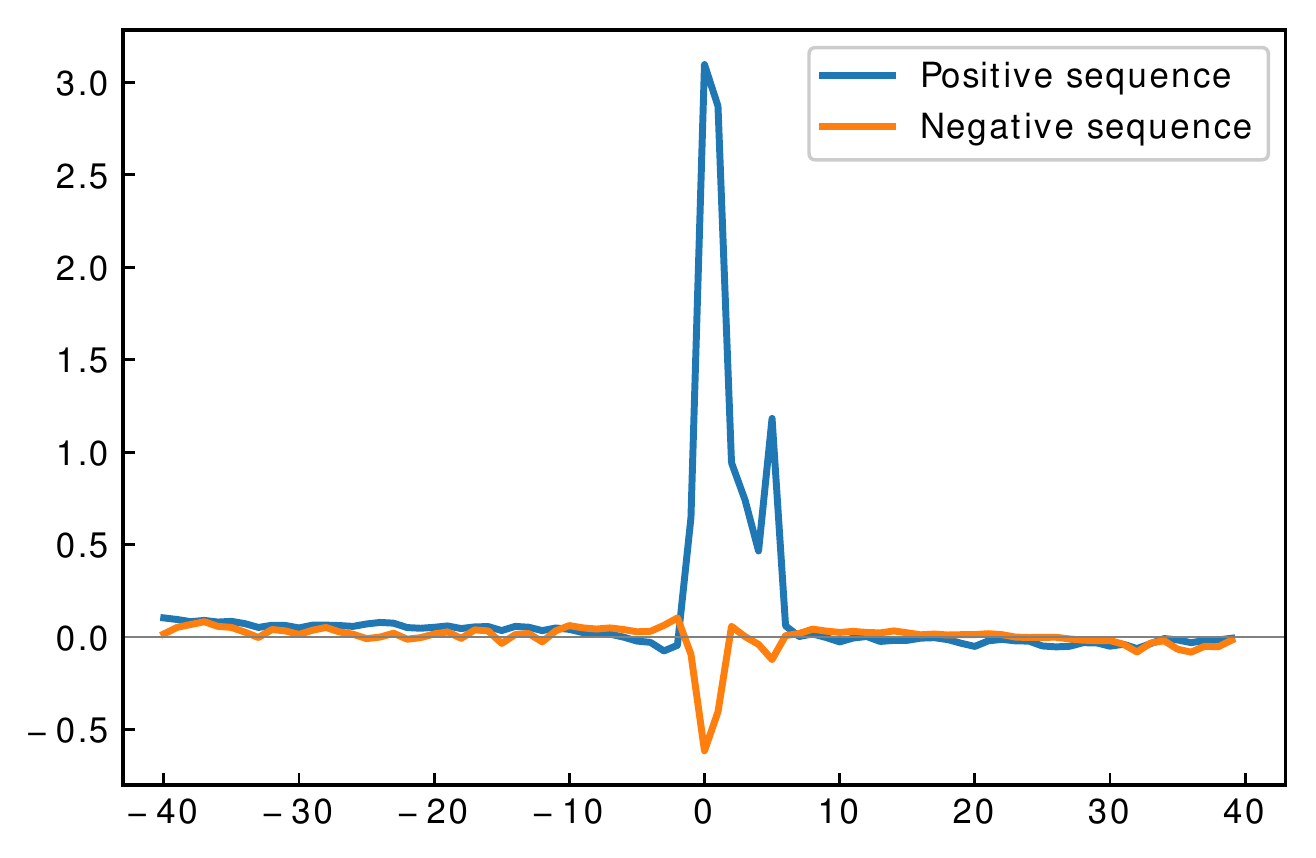}}
\subfigure[Acceptor]{\label{pmacc}\includegraphics[width=40mm]{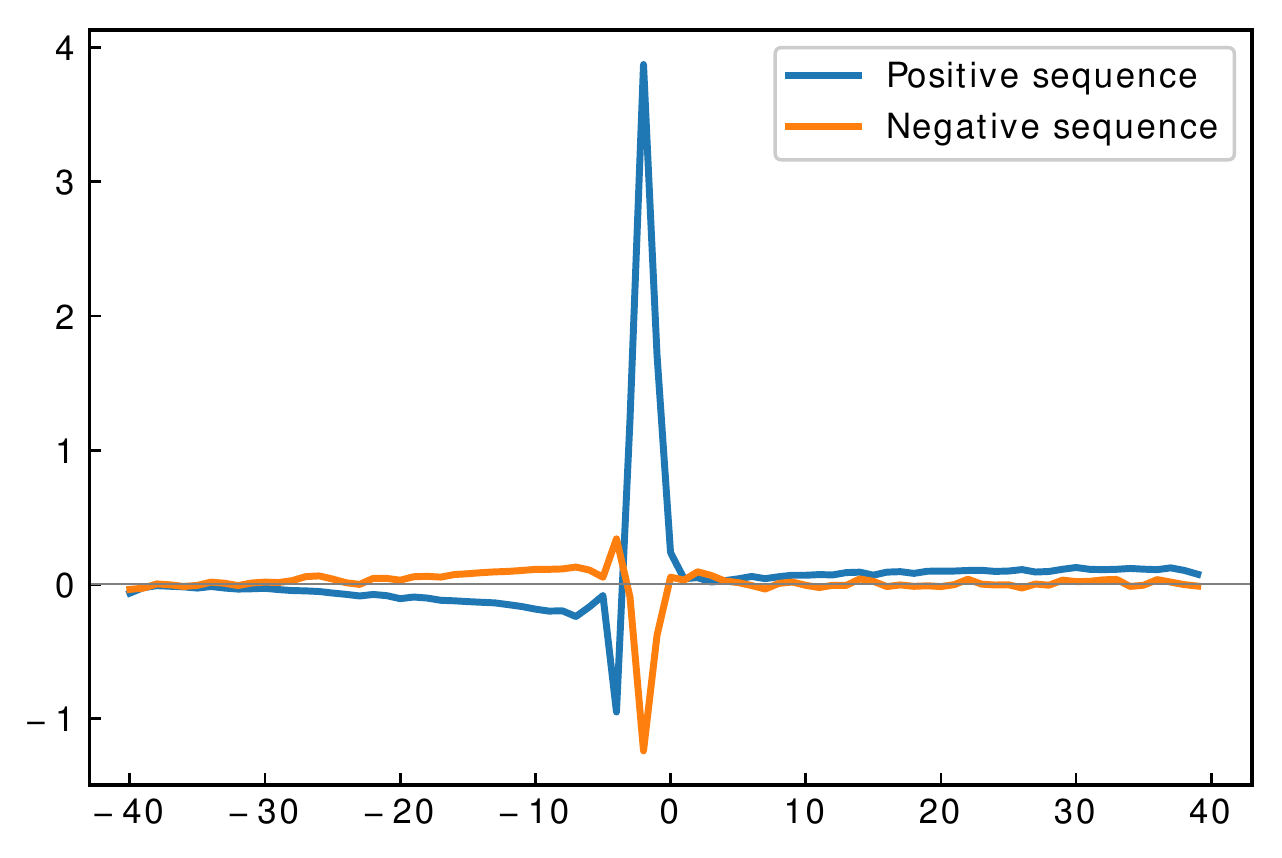}}
\caption{Average attribution score per position for positive and negative sequences}
\end{center}
\vskip -0.3in
\end{figure}

\begin{figure}[t]
\vskip 0.2in
\begin{center}
\label{totalnscore}
\subfigure[Donor]{\label{totalndon}\includegraphics[width=40mm]{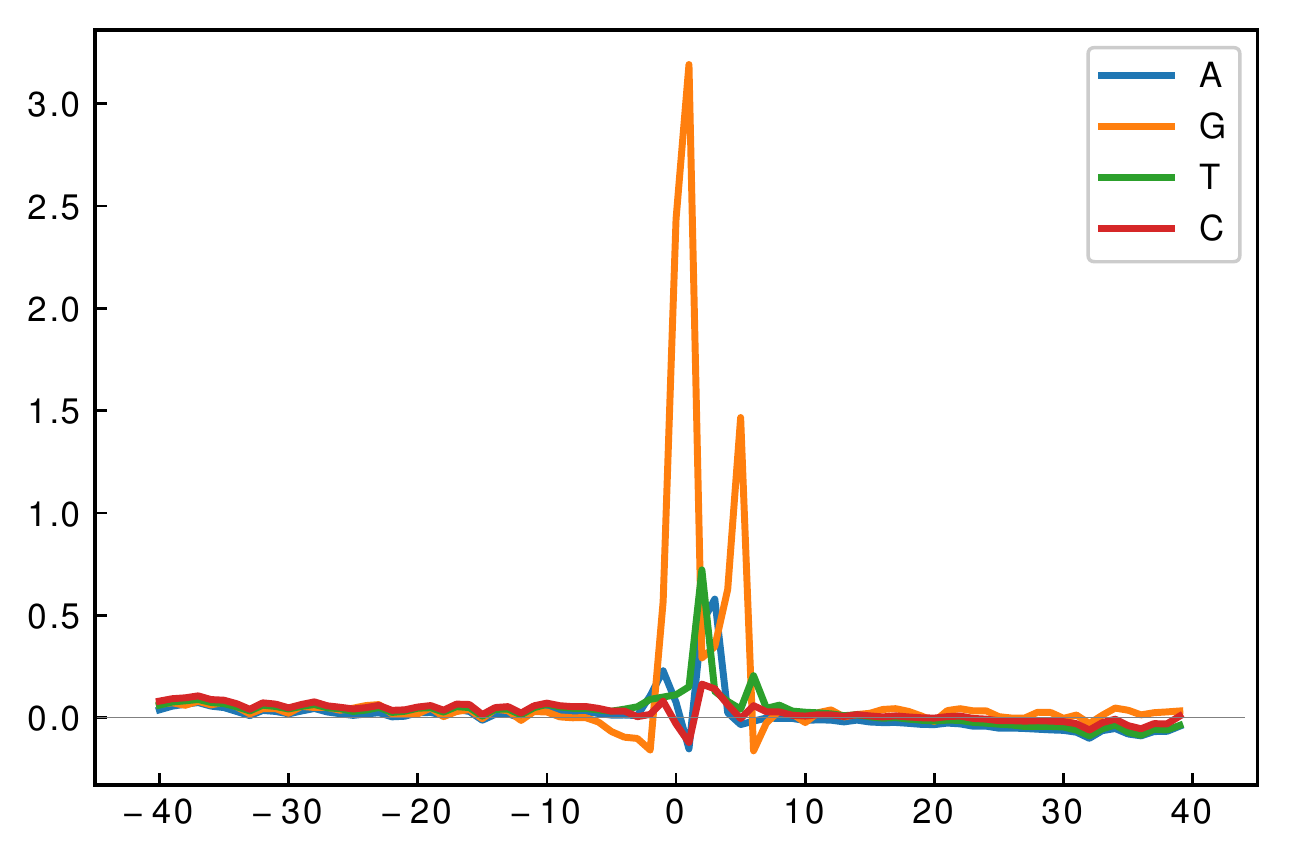}}
\subfigure[Acceptor]{\label{totalnacc}\includegraphics[width=40mm]{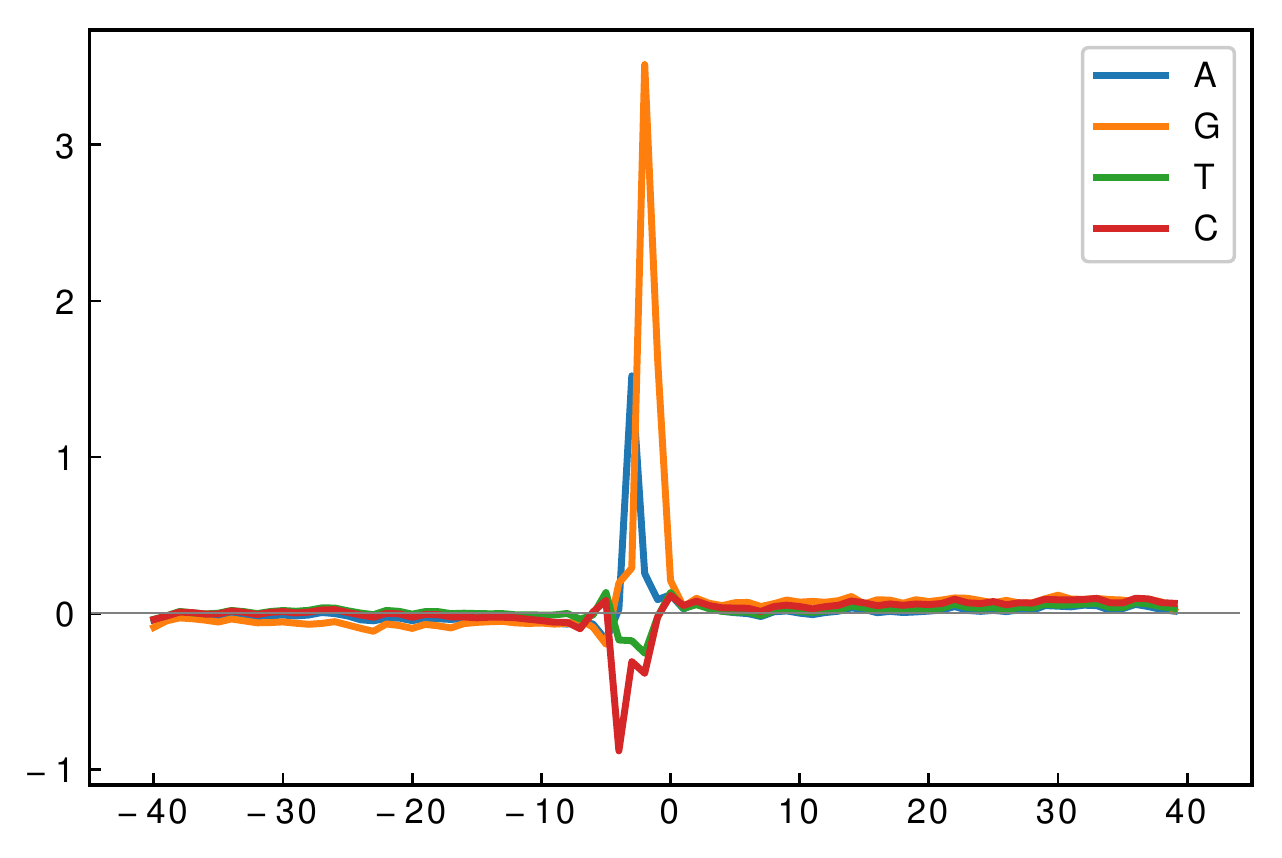}}
\caption{Average attribution score per position for each nucleotide}
\end{center}
\vskip -0.3in
\end{figure}

\begin{figure}[t]
\vskip 0.2in
\begin{center}
\subfigure[Donor]{\label{logodon}\includegraphics[width=40mm]{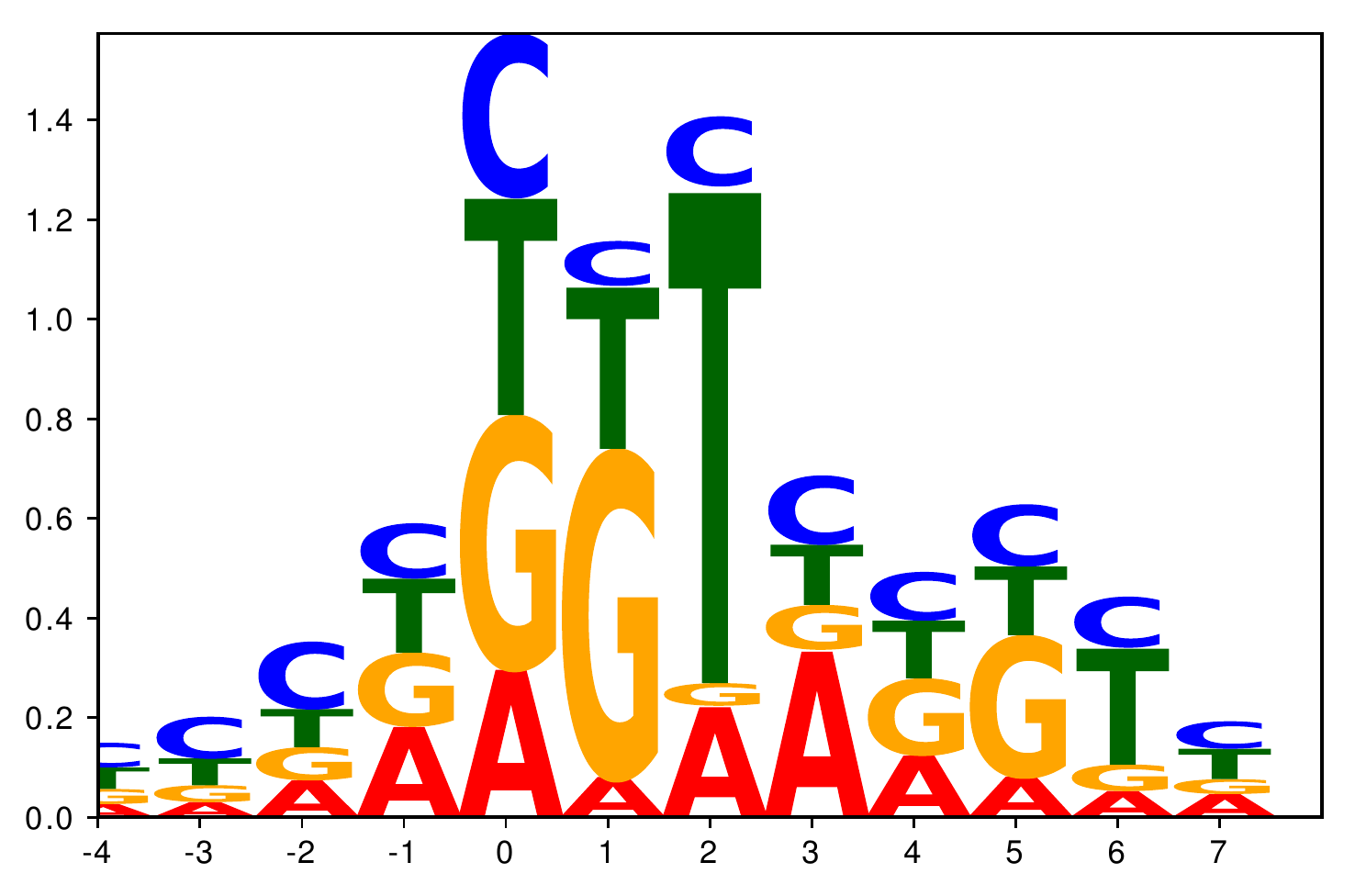}}
\subfigure[Acceptor]{\label{logoacc}\includegraphics[width=40mm]{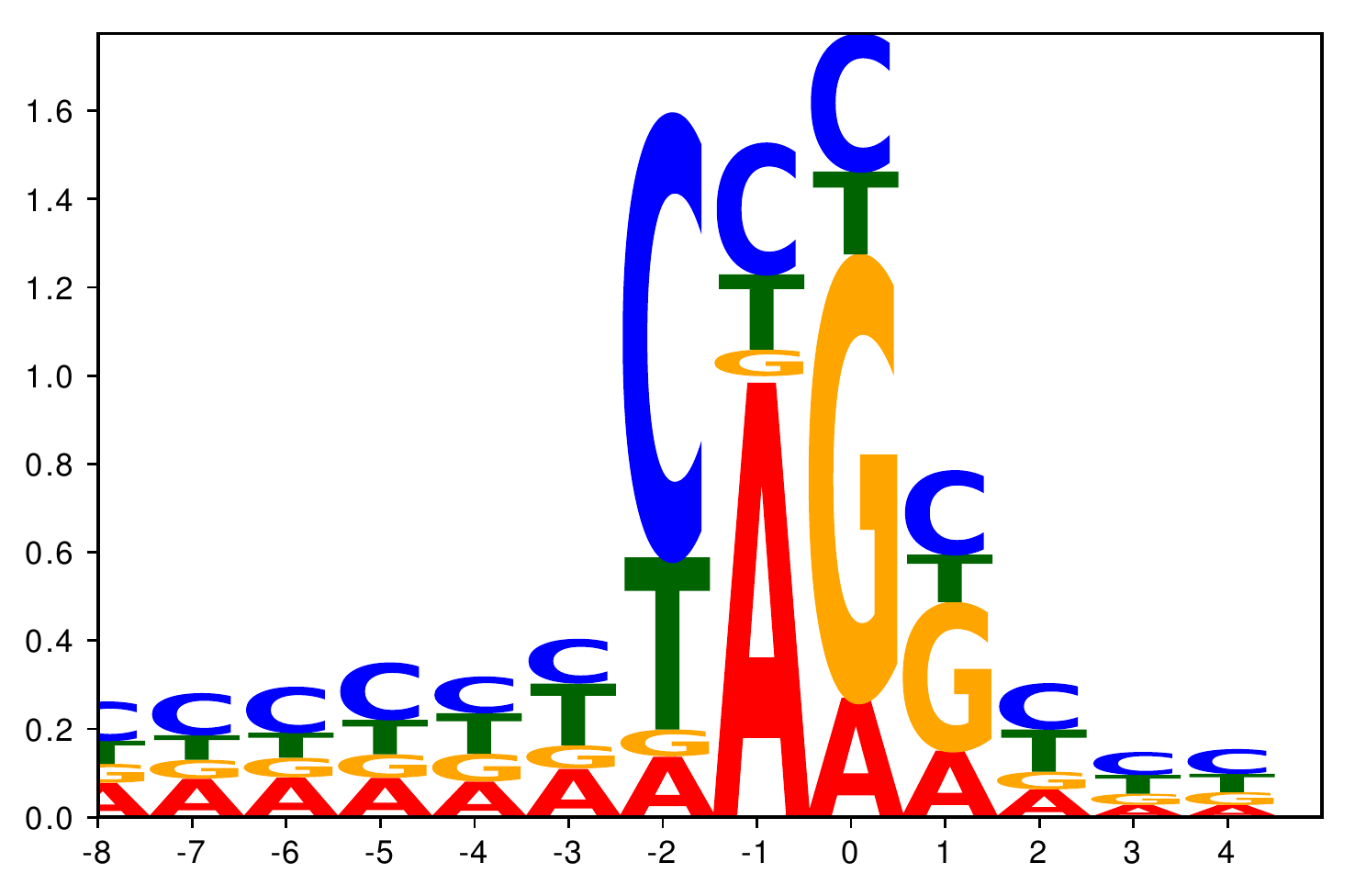}}
\caption{Sequence logo to visualize important motifs attributed by the model}
\label{seqlogo}
\end{center}
\vskip -0.3in
\end{figure}

For visualization, we take 40nt sequence window upstream and downstream around both at acceptor and donor sites. 
Figure \ref{totaldon} and \ref{totalacc} shows the attribution score averaged over all data for donor and acceptor respectively. It is evident that the nucleotides closer to the sites influenced the model most, which confirms with the existing knowledge. The similar analysis for positive and negative example sequences are shown in Figure \ref{pmdon} and \ref{pmacc}. We also perform analysis of attribution score for all nucleotide A, T, G and C separately and the comparison result for both donor and acceptor sites is shown in Figure \ref{totalndon} and \ref{totalnacc}. In this case also, the observed result confirms with the existing knowledge. For example, the nucleotide G is very important for the donor site. Finally, we generate a sequence logo from the attribution score to identify important motifs or splicing signals. Figure \ref{logodon} and \ref{logoacc} shows the sequence logo for both donor and acceptor around the most relevant region given by the attribution score. The donor results show the presence of strong GT signal and the acceptor results shows the presence of strong AG signal. This validates the known consensus motif.

\section{Conclusion}

In this work, we presented an unsupervised representation learning approach to learn representations of DNA sequences in a latent space. We leveraged deep learning techniques to use a sequence-to-sequence autoencoder-like framework to learn representations in an unsupervised setting. We exploit this autoencoder model in two ways: first the learned weight parameters of this model can be used to initialize a classifier with similar architecture, and second, latent representation was used as input features for three different classifiers SVM, ANN and vanilla RNN. The results indicate that the use of pre-trained weight parameters help in faster convergence with improved accuracy. Furthermore, our analysis shows that the learned latent embeddings are good features as three different classifiers gave similar performance using it as input feature. Finally, attributional analysis shows that the model is able to pick significant regions, confirming with the existing knowledge, of input DNA sequence for the splice site classification task.



\nocite{srivastava}
\bibliography{example_paper}
\bibliographystyle{icml2019}

\end{document}